\documentclass[apj]{emulateapj}
\usepackage{mathptmx}
\journalinfo{The Astrophysical Journal, 
???: ??1--??7, 2003 Month, astro-ph/0305200}
\slugcomment{Received 2003 May 6; accepted 2003 May 9}

\def\gtorder{\mathrel{\raise.3ex\hbox{$>$}\mkern-14mu
             \lower0.6ex\hbox{$\sim$}}}
\def\ltorder{\mathrel{\raise.3ex\hbox{$<$}\mkern-14mu
             \lower0.6ex\hbox{$\sim$}}}

\shorttitle{TIME DELAY MEASUREMENT OF HE1104$-$1805}

\shortauthors{OFEK \& MAOZ}

\begin{document}

\title{Time delay measurement of the lensed quasar HE1104$-$1805}
\author{
Eran~O.~Ofek and Dan~Maoz
}
\affil{School of Physics and Astronomy, and Wise Observatory, Tel-Aviv University, Tel-Aviv 69978, Israel.}
\email{eran@wise.tau.ac.il}

\begin{abstract}
We have measured the time delay between the two images
of the gravitationally lensed quasar HE1104$-$1805
by combining observations made with the Wise Observatory 1m telescope
and published observations of this system by Schechter et al.,
spanning a total of five years, from 1997 to 2002.
Based on a variety of techniques, we find that
the best fit time delay is $-161_{-7,-11}^{+7,+34}$~days.
The $68\%$ and $95\%$ confidence intervals include the systematic errors
due to an observed component of uncorrelated variability
between images.
The delay is shorter than predicted by
simple models and may indicate a large external shear
or a large value of the Hubble parameter, $h>0.75$ ($95\%$ confidence).
The residual light curve between the two images shows
a longterm trend of $\sim0.04$~mag~yr$^{-1}$,
presumably the result of microlensing by stars in the lens galaxy,
but also short timescale ($\sim1$~month) variability
with a mean amplitude of about $0.07$~mag.

\end{abstract}


\section{Introduction}
\label{Introduction}

Time-delay measurements in a large sample
of gravitationally lensed systems could be used to measure
the Hubble parameter, $H_{0}$,
directly at high redshift, therefore avoiding
the systematics due to distance ladder calibrations
and large-scale motions (e.g., Turner, Cen, \& Ostriker 1992).
Alternatively, using a measured value
of the Hubble parameter from other techniques (e.g., Bennett et al. 2003),
measurement of a time delay allows
discriminating between different
mass profiles of galaxies (e.g., Kochanek 2002).
In some systems,
complex geometries caused by multiple lensing
galaxies hinder the use of time delay measurements
for $H_{0}$ determination or for mass profile studies.
Even in such systems,
time delay measurements are important for isolating
the microlensing variability
from the intrinsic variability.
Microlensing in lensed quasars could allow
an independent measurement of the fraction of dark matter
in galaxies (Schechter \& Wambsganss 2002),
and can be used to put limits on the size of the continuum emission regions in
quasars (e.g., Wambsganss, Paczynski, \& Schneider 1990; Wambsganss 2002;
Wyithe, Agol, \& Fluke 2002).

HE1104$-$1805 is a double-image lensed quasar
discovered by Wisotzki et al. (1993).
The image separation is $\Delta{\theta}=3.19''$,
the source redshift is $z_{s}=2.319$ (Wisotzki et al. 1993),
and the lens redshift is $z_{l}=0.729$ (Lidman et al. 2000).
This is an unusual system, in which the image
closer to the lens (image $A$) is the brighter one.
Wisotzki et al. (1993) noted the emission line flux ratio between the images
is $2.8$, and that component $A$ has a bluer continuum
and lower equivalent width emission lines than $B$.
They interpreted the differences as being caused by 
microlensing of the quasar continuum source in the brighter image.
Wisotzki et al. (1995) reported that the continuum flux in both
images is highly variable, but that the line fluxes
do not change. Again, this could be interpreted as evidence
for strong microlensing.
Leh{\' a}r et al. (2000) have attempted modeling this system.
They assumed an image flux ratio of about 4, and found
this limits the models to a narrow range of predicted
time delays.

Wisotzki et al. (1998) measured the continuum flux
in $19$ low resolution spectra, taken with the
ESO 3.6m telescope between 1993 to 1998.
By matching the light curves
of the two images, they favored a time delay of about
$-270$~days (in the sense that the $B$ image leads the $A$ image),
although they concluded that a value as short
as $-100$~days could not be excluded.
Gil-Merino, Wisotzki, \& Wambsganss (2002)
analyzed the $B$-band acquisition frames for the spectroscopy
of Wisotzki et al. (1998) and found
a best fit time delay of $-310\pm20$~days ($2\sigma$ errors).
Pelt, Refsdal, \& Stabell (2002) re-analyzed the Gil-Merino et al. (2002)
observations, and argued that
the time delay is somewhere between $-330$ and $-255$~days,
but with a large uncertainty.
In a recent paper,
Schechter et al. (2003) presented 3 years of $V$-band
photometry of HE1104$-$1805,
obtained on $102$ nights with the OGLE 1.3m telescope.
They did not succeed in finding a consistent time delay for the system.
In fact, the root-mean-square (rms) of the difference
between the component light curves as a function of time delay
showed two minima, at about $-150$~days and $-360$~days.
Moreover, Schechter et al. (2003) noted that the
structure functions of images $A$ and $B$ are quite different,
with image $A$ being more than twice as variable as image $B$
on timescales of less than a month.
Thus, there is presently no clear evidence of a
time delay in this system,
let alone an unambiguous measurement of its value.

In this paper we present new photometric data
for HE1104$-$1805 from the Wise Observatory
lens monitoring project.
Combined with the OGLE observations
of HE1104$-$1805 from Schechter et al. (2003),
our data establish the delay securely
and isolate the various correlated and uncorrelated variability
components.
We will assume throughout
a cosmology with $\Omega_{m}=0.3$, $\Omega_{\Lambda}=0.7$,
and $H_{0}=100h$~km~s$^{-1}$~Mpc$^{-1}$.

\section{Observations and Reduction}
\label{Observations}

For the past three years we have been monitoring a sample of
$27$ known gravitationally lensed quasars and lens candidates
with the Wise Observatory 1m telescope.
The objects are monitored on a weekly basis in the
Johnson-Cousins $R$-band,
with occasional coverage in the $I$-band or $V$-band.
Frames are obtained with a cryogenically cooled
Tektronix $1024\times1024$-pixel back-illuminated CCD.
The scale is $0.7''$ per pixel.
The median seeing of
about $2''$ does not allow resolving most of the lensed objects.
However, some of the pairs are resolved
(e.g., Q0142$-$100, HE1104$-$1805, RXJ0921$+$4528, Q0957$+$561).
The data are reduced automatically on a
daily basis.
In each field we perform aperture photometry
of all sources in the frame and the preliminary combined-image light curves are
posted on the WWW\footnote{http://wise-obs.tau.ac.il/$\sim$eran/LM/}.

We present here $R$-band measurements of
HE1104$-$1805 taken between 1999, November 14 and 2002, June 18.
The combined OGLE and Wise observations span over five years,
from 1997 to 2002.
The Wise observatory photometry of HE1104$-$1805 was performed using
a special program we have written for this purpose.
All the frames are aligned and rotated to the same
reference frame.
The program constructs a numerical point-spread function (PSF) from
preselected reference stars.
Given accurate positions of the lensed images
from Hubble Space Telescope
images (Leh{\' a}r et al. 2000), it fits the PSF
to the images and minimizes the $\chi^{2}$
(assuming Poisson noise) with respect to the two free
parameters, the magnitude
of the two images, $m_{1}$ and $m_{2}$.
In the case of HE1104$-$1805, the PSF was constructed from $12$ stars.
After rejecting all measurements with errors larger than $0.1$~mag,
we are left with $79$ epochs for image $A$ and $49$
epochs for image $B$\footnote{The measurements, as well as finding chart with reference stars, are available from http://wise-obs.tau.ac.il/$\sim$eran/LM/HE1104/}.
Note that the light contributed by the lensing galaxy
to image $A$
was neglected in the photometry, as it contributes only $0.3\%$ and $3.5\%$
of the total light of image $A$ in the $V$ and $I$ bands, respectively.

The error bars were calculated from the 
covariance matrix of the $\chi^{2}$ surface
(e.g., Press et al. 1992).
To validate the reliability of the error bars, we have
made an empirical, magnitude dependent, error estimate
for each lensed image and compared it with
the $\chi^{2}$ error estimate.
The empirical errors were calculated as follows.
Using the same algorithm,
we have measured
the difference, $\Delta{m}_{i}^{t}$, for each frame taken at epoch $t$,
between the magnitude of
each reference star, $i$, and its average magnitude over time, $\bar{m}_{i}$
(with the magnitudes measured relative to the $12$ co-added reference stars).
The error in the measurement of a particular source will
depend on its intrinsic magnitude and on the observing conditions on
the particular epoch.
To find this dependence,
we fit a function, with two free parameters $p$ and $q$, of the form:
$\log{\Delta{m}_{i}^{t}} = p + q \bar{m}_{i}$ to all the frames,
and find $q=0.3$.
This function approximately describes the expected dependence
of error on magnitude.
Note that,
in the case that the background Poisson noise is
the dominant source of error, we expect $q=0.4$,
while in the limit of zero background we expect $q=0.2$.
Assuming that $q=0.3$ is constant over all nights,
for each individual frame we refit this formula
to obtain $p_{t}$.
This fit gives, for each frame,
the typical error as a function of magnitude.
Finally, for each frame, we have compared the $\chi^{2}$ errors
in the magnitudes
of the lensed images, $A$ and $B$ ($m_{t}^{A}$, $m_{t}^{B}$),
with the empirical errors $10^{p_{t} +qm_{t}^{A}}$ and $10^{p_{t} +qm_{t}^{B}}$.
We find excellent agreement between the two error estimators.

Table~1 lists the reference star coordinates, magnitudes, $\bar{m}_{i}$,
and the $68\%$ confidence interval of their magnitude distribution ($68\%$~CI).
All magnitudes in this paper are given relative to reference star $S7$.
\begin{deluxetable}{lcccc}
\tablecolumns{5}
\tablewidth{0pt}
\tablecaption{PSF reference stars}
\tablehead{
\colhead{Ref} &
\multicolumn{2}{c}{R.A. (J2000) Dec.} &
\colhead{$\bar{m}_{i}$} &
\colhead{$68\%$CI}
}
\startdata
$S1$   & 11:06:33.17  & $-$18:21:39.1 & $0.262$     & $0.025$ \\
$S2$   & 11:06:33.86  & $-$18:20:29.6 & $0.585$     & $0.038$ \\
$S3$   & 11:06:44.75  & $-$18:20:43.2 & $0.374$     & $0.035$ \\
$S4$   & 11:06:38.06  & $-$18:19:10.7 & $0.344$     & $0.043$ \\
$S5$   & 11:06:30.38  & $-$18:19:33.0 & $1.633$     & $0.088$ \\
$S6$   & 11:06:29.74  & $-$18:20:02.3 & $1.321$     & $0.065$ \\
$S7$   & 11:06:23.23  & $-$18:19:41.5 & $0.000$     & $0.028$ \\
$S8$   & 11:06:22.50  & $-$18:21:30.4 & $0.384$     & $0.038$ \\
$S9$   & 11:06:35.63  & $-$18:21:29.3 & $0.987$     & $0.055$ \\
$S10$  & 11:06:37.00  & $-$18:22:45.5 & $0.414$     & $0.033$ \\
$S11$  & 11:06:37.39  & $-$18:23:00.3 & $1.551$     & $0.100$ \\
$S12$  & 11:06:38.31  & $-$18:22:48.1 & $1.284$     & $0.060$ \\
\enddata
\tablecomments{Astrometry is based on $38$
USNO-A2.0 stars (Monet 1998) with $0.''4$~rms in each axis.
Magnitudes are given relative to reference star $S7$.}
\end{deluxetable}

\section{Wise-OGLE Intercalibration}
\label{intercalib}

We have intercalibrated the OGLE and Wise photometry,
in order to produce combined light curves with a long time coverage.
To this end, we take advantage of the fact that the Wise and OGLE
light curves overlap in the year 2000 (J.D. $\sim2451450$-$2451750$).
However, the OGLE observations were obtained in the $V$ band,
while the Wise observations were obtained in the $R$ band.
Wisotzki et al. (1995) have already noted that
the flux ratio between the quasar images depends on wavelength.
We therefore intercalibrated the photometry
for the light curve of each lensed image separately.
For each image, we performed a $\chi^{2}$
fit between the overlapping OGLE and Wise light curves
to determine the magnitude offset and an
optional stretch factor.
We allow for a stretch factor since quasars are
known to vary in color, and the amplitude of variation
in the different bands is thus not necessarily the same.
We used two types of fits:
(i) An offset + stretch factor, $m_{R}=a+bm_{V}$, where
$m_{R}$ and $m_{V}$ are the magnitudes in the $R$ and $V$ bands, respectively;
and (ii) A pure magnitude offset, $m_{R}=a+m_{V}$.

Since the OGLE and Wise observations
during the overlap period were not carried out
on the same days, we linearly interpolated the OGLE observations
(which are more frequent and have smaller error bars)
to the times of the Wise observations.
We set the error of each interpolated magnitude to be
\begin{equation}
\epsilon = \sqrt{\epsilon_{n}^{2} + sf(\Delta{t}_{min})},
\label{sf_prop_err}
\end{equation}
where $\epsilon_{n}$ is the error of the real measurement nearest in time
to the interpolated point,
$sf(\Delta{t}_{min})$ is the value of the structure function
of the interpolated light curve
at lag $\Delta{t}_{min}$,
and $\Delta{t}_{min}$ is
the time between the interpolated epoch and the nearest
epoch which has a real measurement.
The structure function is defined as
\begin{equation}
sf^{2}(t) = 2 [V - DCF(t)],
\label{sf}
\end{equation}
where $V$ is the variance of the light curve and $DCF(t)$ is the
(unnormalized) discrete auto-correlation function (Edelson \& Krolik 1988),
calculated with $14$-day bins.
This scheme gives the interpolated points a realistic weight
in the $\chi^{2}$ minimization.

The $1\sigma$ errors of the
pure magnitude-shift fit between the $V$ and $R$ light curves
are less than $0.03$~mag for both images.
Applying the shift+stretch fit,
the best-fit stretch factor between the $V$ and $R$ light curves
is consistent with unity
and does not improve the fit significantly.
As we will show below,
allowing for a non-unity stretch factor does not
change the best fit time delay.
Therefore, in what follows 
we will use the magnitude-shifted light curves (i.e., with no stretch).
We test in \S\ref{TimeDelay} for the
uncertainty in the time delay induced by the range of
possible shift+stretch parameters.

\begin{figure}
\centerline{\includegraphics[width=8.5cm]{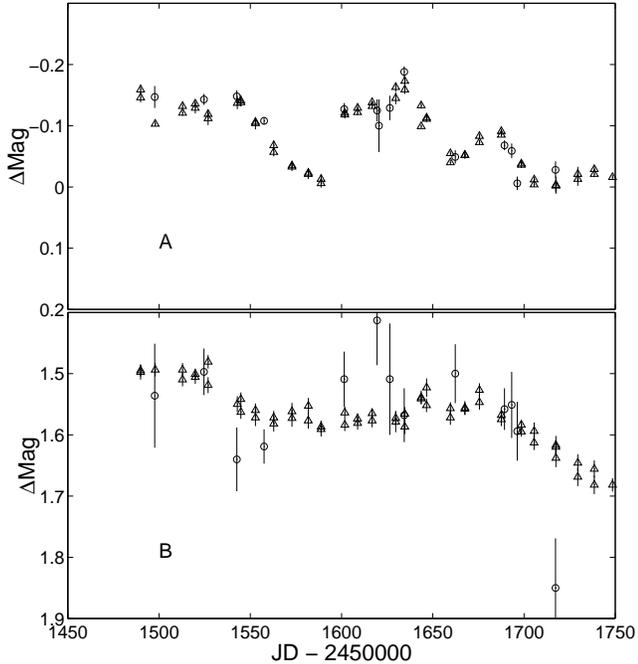}}
\caption{Image $A$ (upper panel) and image $B$ (lower panel)
light curves during the period of overlap between
the $V$-band OGLE observations (triangles) and the $R$-band Wise observations (circles).
The observations were brought to the same scale by
$\chi^{2}$ minimization.
Magnitudes are relative to reference star $S7$.
\label{ZoomOnFitLC} }
\end{figure}

\begin{figure*}[t]
\centerline{\includegraphics[width=18.0cm]{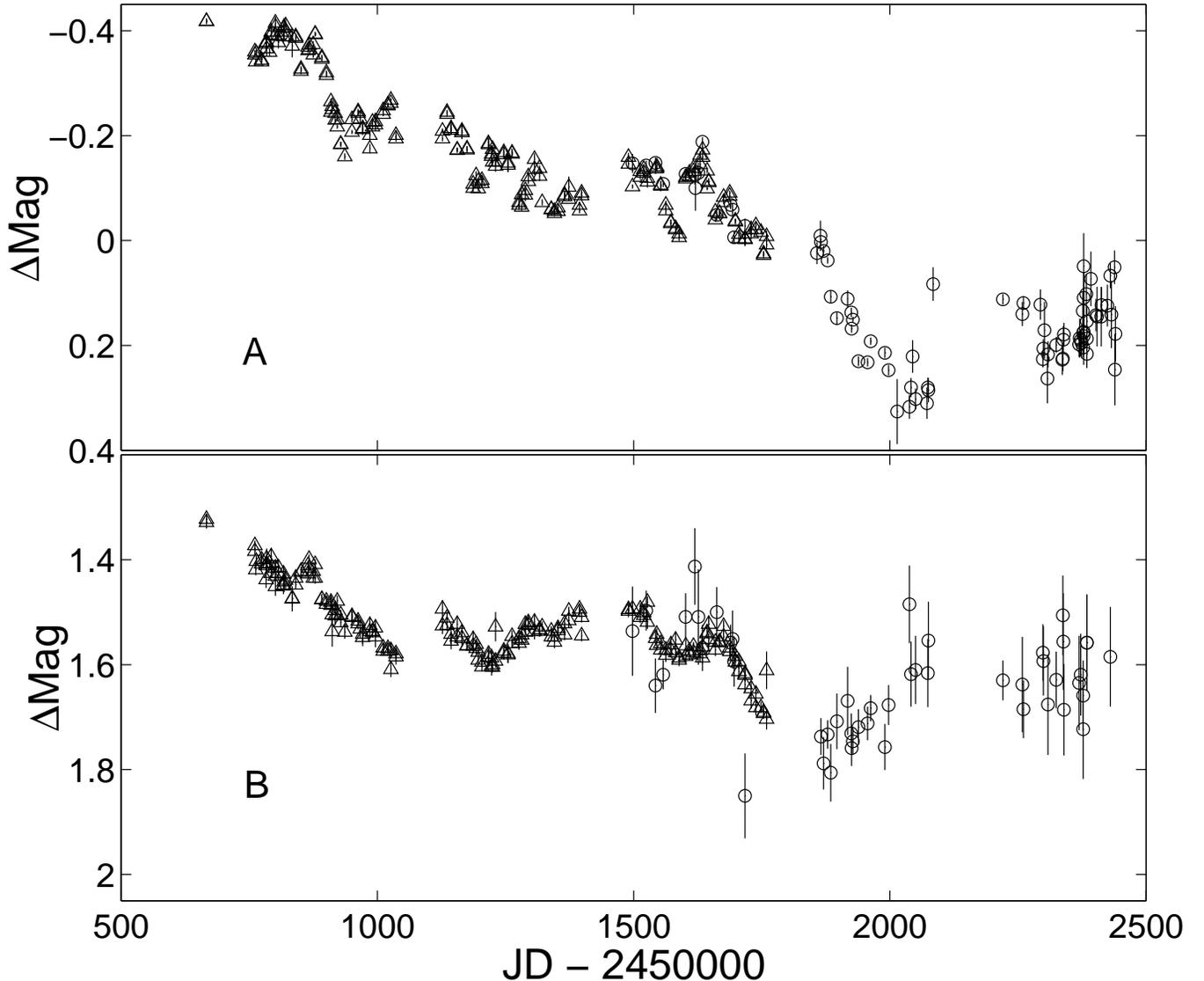}}
\caption {Image $A$ (upper panel) and image $B$ (lower panel)
light curves from 1997 to 2002, based on
the $V$-band OGLE observations (triangles) and the $R$-band
Wise observations (circles),
after intercalibration of the two data sets.
\label{CombinedLC} }
\end{figure*}

The combined light curves in the zone of overlap are shown in
Figure~\ref{ZoomOnFitLC},
for images $A$ (upper panel) and $B$ (lower panel).
In Figure ~\ref{CombinedLC} we show the
complete (1997-2002) light curves
of images $A$ (upper panel) and $B$ (lower panel),
after application of the intercalibration to the whole dataset.
The complete intercalibrated data,
from Schechter et al. (2003) and our observations,
are given in Table~2.
\begin{deluxetable}{llll}
\tablecolumns{4}
\tablewidth{0pt}
\tablecaption{Intercalibrated image A light curve}
\tablehead{
\colhead{JD-2450000} &
\colhead{mag} &
\colhead{error} &
\colhead{Observatory\tablenotemark{a}}
}
\startdata
$666.46490$ &  $-0.418$ &  $0.005$ & $1$  \\
$666.47390$ &  $-0.418$ &  $0.005$ & $1$  \\
$760.84802$ &  $-0.360$ &  $0.003$ & $1$  \\
$760.85665$ &  $-0.354$ &  $0.003$ & $1$  \\
$762.85473$ &  $-0.341$ &  $0.004$ & $1$  \\
\enddata
\tablenotetext{a}{1- Observations from Schechter et al. (2003; OGLE V band); 2- Our observations (Wise R band).}
\tablecomments{Table~2a in its entirety, as well as Table~2b for image B, are available via the electronic version.}
\end{deluxetable}

\section{Time Delay between Images A and B}
\label{TimeDelay}

We now search for the time delay between images $A$ and $B$
by using $\chi^{2}$ minimization.
From Figure~\ref{CombinedLC}, it is apparent that the long term variation
of image $A$ is more pronounced than that of image $B$.
This could be the result of microlensing in the system.
Thus, in the $\chi^{2}$ minimization
we leave as free parameters not only a time delay,
but also a linear trend of magnitude with time
between the light curves.
The fit is described by
\begin{equation}
\chi^{2}=\sum_{i}^{N}{\frac{[m_{t+\tau}^{B} - m_{t}^{A} + S(t-t_{mid}) - C]^2
}{\epsilon_{n}^{2} + sf(\Delta{t}_{min})}}
\label{FitEquation}
\end{equation}
where $m_{t}^{A}$ and $m_{t}^{B}$ are
the magnitudes as a function of the time, $t$, for images $A$
and $B$, respectively.
The arbitrary constant $t_{mid}$ is defined as the midpoint between
the first and last observations (i.e., $JD=2451525.378$),
$C$ is the mean
magnitude difference between the images,
and $S$ and $\tau$ are the fit parameters -- a linear trend
between the light curves, and a time delay, respectively.   
The fit was performed,
simultaneously for $S$ and $C$,
between the
light curve of image $A$ and each of the $\tau$-shifted
light curves of image $B$.
In this process, the light curve of image $A$ at epoch $t$
was interpolated
to the times ($t+\tau$) of the image $B$ observations.
Again, we used linear interpolation and
set the error of the interpolated magnitude
using Eq.~\ref{sf_prop_err}.
%
%
The bin size used for calculating the DCF
necessary for determining the structure function, $sf(\Delta{t}_{min})$,
has a small effect ($<2$~days) on the best-fit time delay.

\begin{figure}
\centerline{\includegraphics[width=8.5cm]{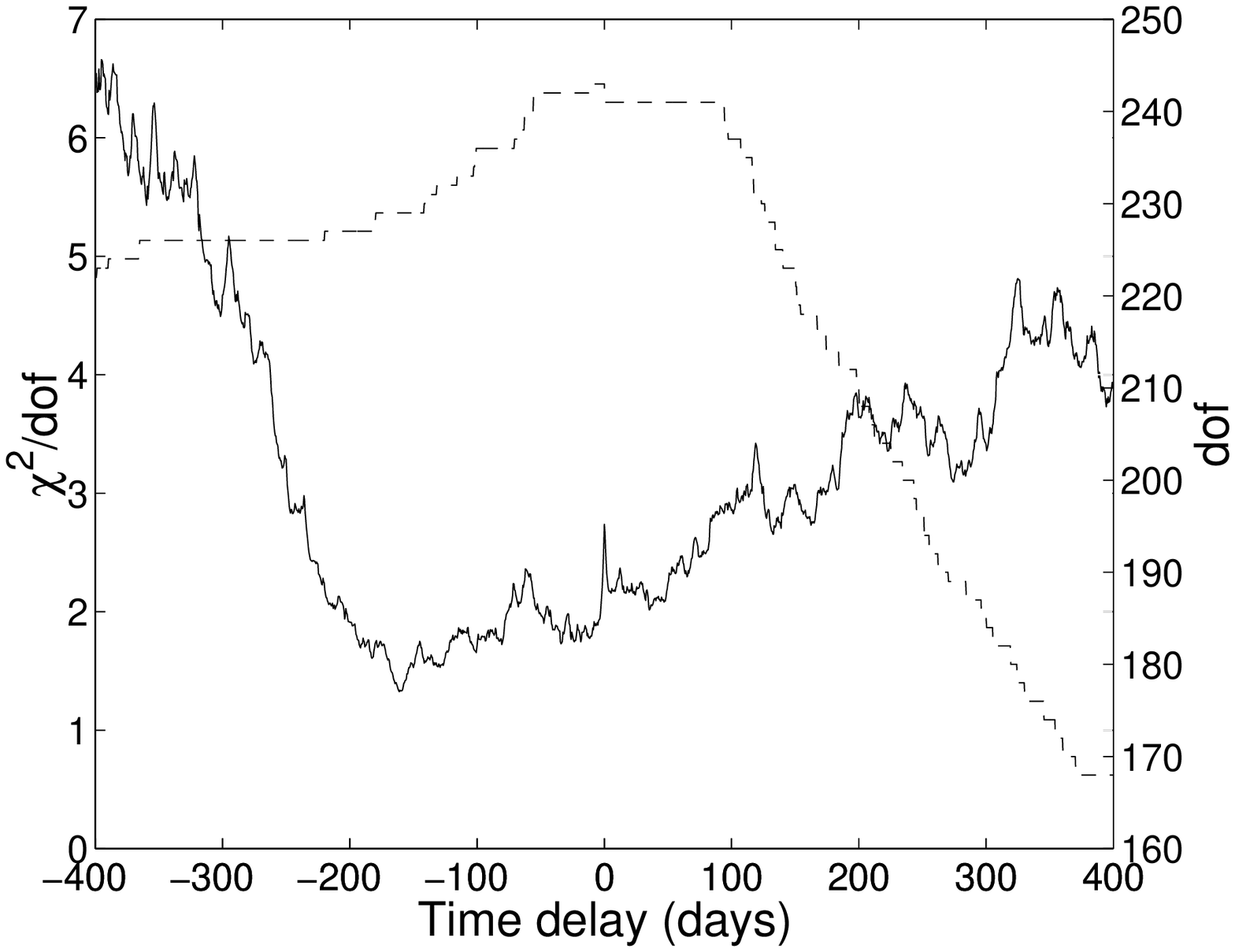}}
\caption {The $\chi^{2}/dof$ as a function of delay time (solid curve).
For each time delay,
the best-fit linear trend ($S$) and magnitude offset ($C$)
were used.
The dashed curve shows the $dof$.
\label{Chi2Fit_SlopeBest} }
\end{figure}

\begin{figure*}
\centerline{\includegraphics[width=18cm]{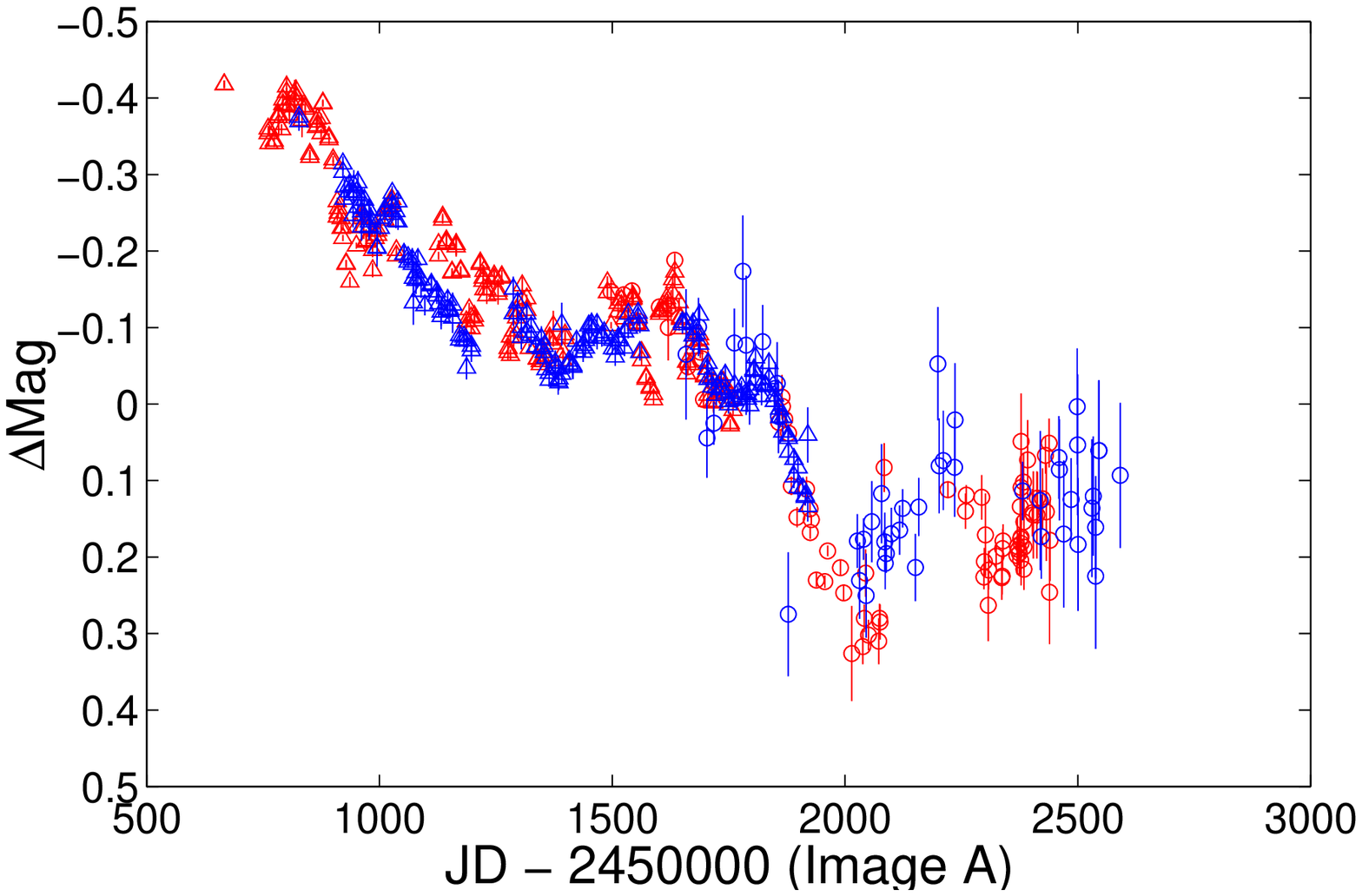}}
\caption {Light curve of image $A$ (red),
overlayed with the best-fitting slope-corrected and time-delay shifted
light curve of image $B$ (blue).
The triangles and circles mark the OGLE and Wise observations, respectively.
\label{ShiftedLC} }
\end{figure*}

Figure~\ref{Chi2Fit_SlopeBest} shows the $\chi^{2}$ per degree of freedom ($dof$)
of the light curve fitting as a function of the time delay.
The dashed curve shows the $dof$.
For each time delay,
the best-fit linear trend ($S$) and magnitude offset ($C$)
were recalculated.
There is a distinct minimum in the $\chi^{2}$ at a time
delay of $-161$~day.
The best fit parameters are
$S=0.043$~mag~yr$^{-1}$ and
$\tau=-160.9$~day (with $\chi^{2}/dof=303/229=1.32$).
For these parameters, the magnitude difference (C) between the images
in the $R$ band is $\Delta{m_{R}}=1.595\pm0.004$~mag.
Figure~\ref{ShiftedLC} shows the light curve of image $A$ (filled circles),
overlayed by the slope corrected ($S=0.043$~mag~yr$^{-1}$)
and time-delay shifted
($\tau=-160.9$~day)
light curve of image $B$ (empty circles).
The formal $\chi^{2}$ errors on the time delay are less than $1\%$.
However, these errors are not realistic.
As we will show below,
there are several sources of systematic errors
that need to be taken into account.

As noted above, a systematic error is introduced
to the time delay by the uncertainty in
the OGLE and Wise intercalibrations
(using the shift algorithm or the shift+stretch algorithm).
We have used the uncertainty in the
shift+stretch parameters and their covariance term,
and have run $1000$ Monte-Carlo simulations in which we
randomly drew shift+stretch parameters from a bivariate Gaussian distribution,
and repeated the time-delay fit.
We find a time delay distribution
of $-159.2_{-1.7,-2.0}^{+0.4,+33}$~day
($68\%$ and $95\%$ confidence errors).
The short time delays, found in some of these simulations,
of about $-130$~day, are
the result of stretch-factor values smaller than $b=0.75$.
Quasars are known to have larger variability amplitudes
at bluer wavelengths, corresponding to $b<1$.
For example, Giveon et al. (1999) found $\Delta(B-R)\approx0.25\Delta{B}$
for a sample of quasars at $z\sim0.2$,
which corresponds to a stretch factor $b=0.75$ between $B$ and $R$.
For HE1104$-$1805, we are observing a smaller wavelength interval,
between $V$ and $R$, but on the other hand,
a higher redshift.
The $V$ and $R$ bands sample the restframe UV, which may have
a stronger dependence of variation amplitude on wavelength.
Given the current uncertainty in the stretch factor,
there is about $4\%$ probability that the
time delay between the images of HE1104$-$1805
is between $-125$ and $-155$~days.
If we limit the stretch parameter to
$0.75<b<1$, then the $95\%$ confidence error on the time delay,
due to the stretch + shift uncertainty,
is smaller than $2$~days.

With these results we can reject the possibility that the time delay
is in the region of $-250$ to $-330$~days, as suggested
by Gil-Merino et al. (2002) and Pelt et al. (2002).
Given the significant microlensing variability observed in this
system (e.g., Schechter et al. 2003),
measuring the time delay from sparsely sampled light curve is
difficult.

\begin{figure}
\centerline{\includegraphics[width=8.5cm]{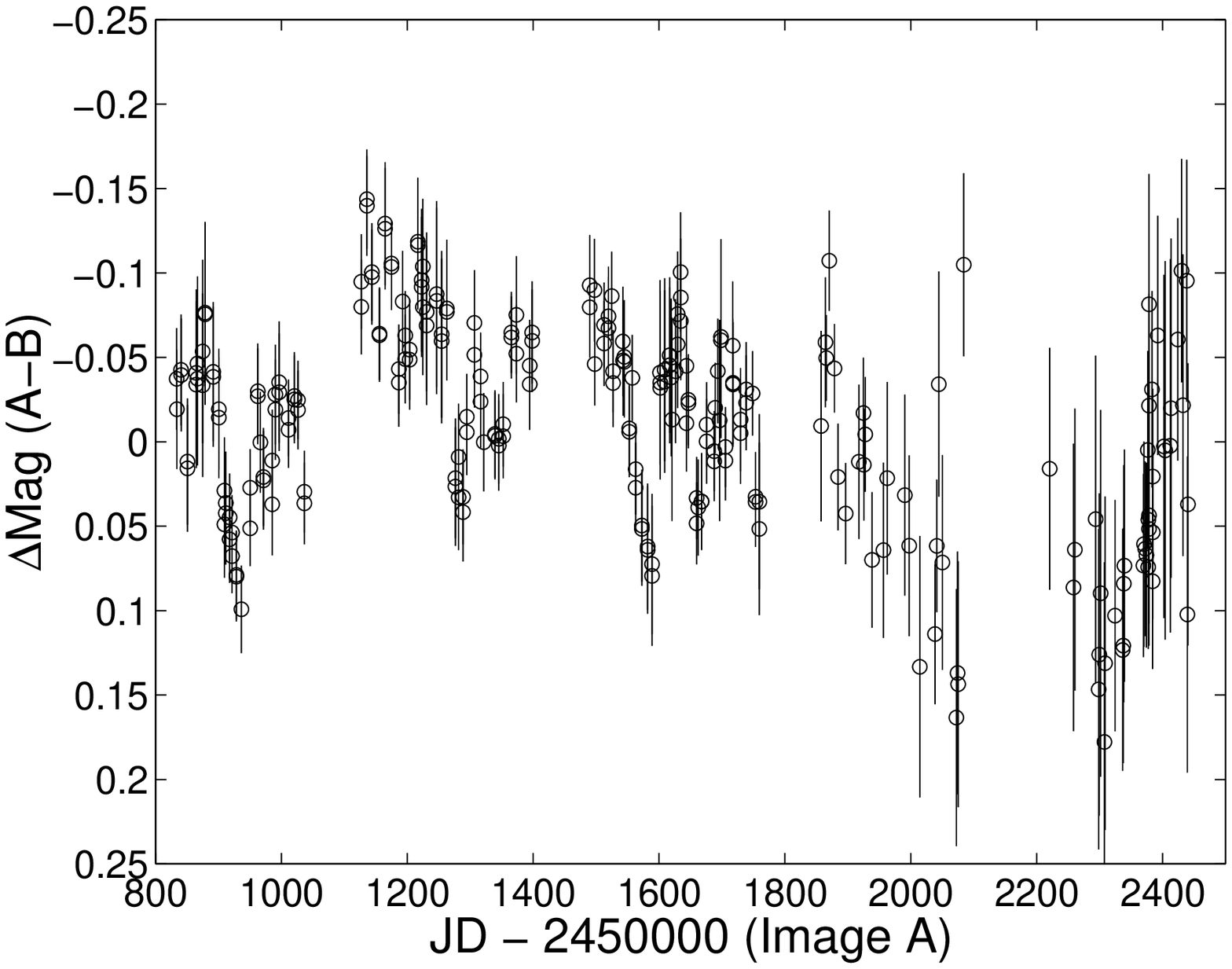}}
\caption {The difference light curve between images $A$ and $B$, after applying
the best-fit linear trend and time delay.
\label{LC1mLC2_Res} }
\end{figure}

\begin{figure}
\centerline{\includegraphics[width=8.5cm]{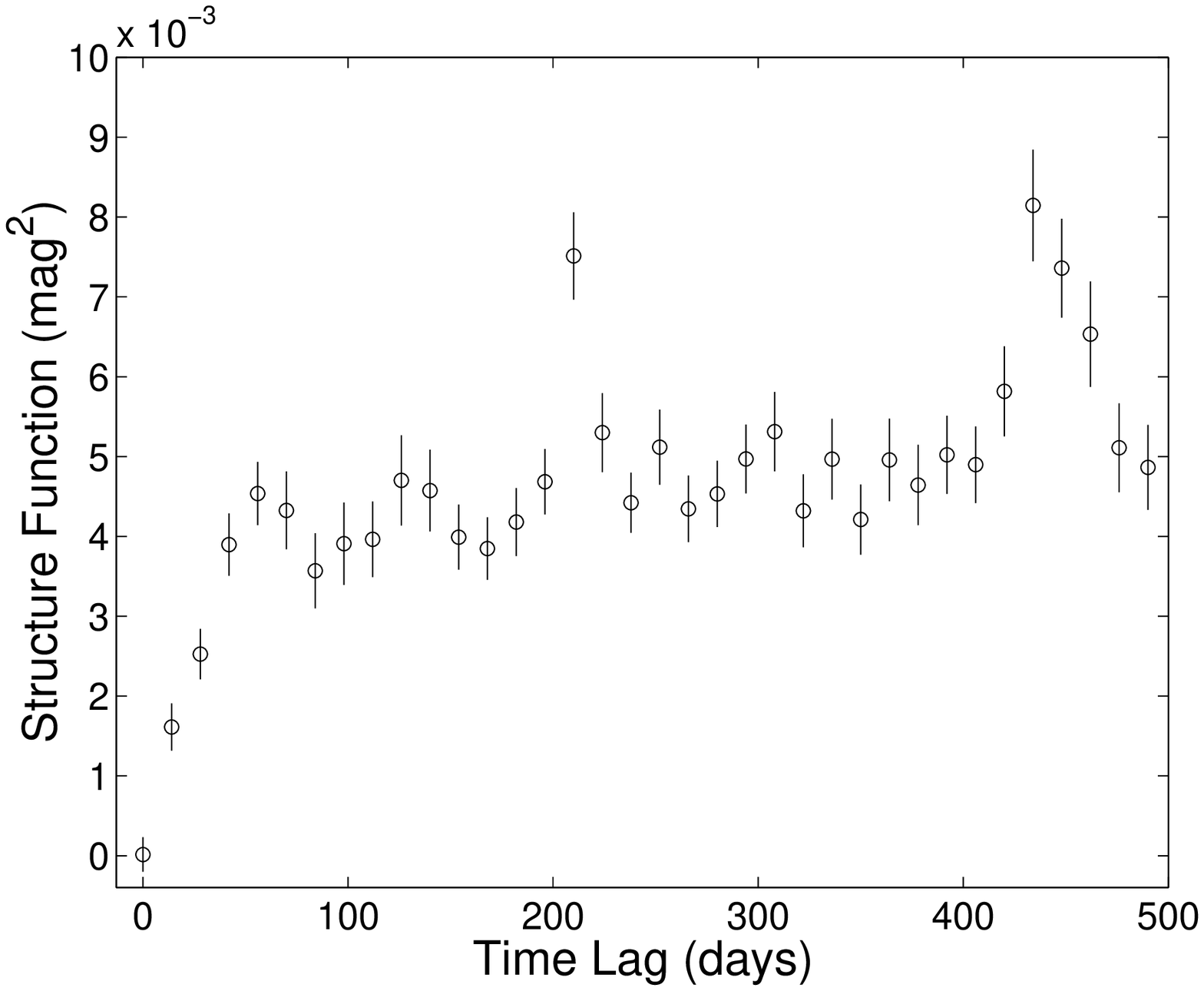}}
\caption {Structure function of the difference light curve of Fig.~\ref{LC1mLC2_Res}.
\label{SF_LC1mLC2_Res} }
\end{figure}

Figure~\ref{ShiftedLC} shows that
the main features and trends appear in
the light curves of both images.
However, there 
are significant coherent fluctuations in the difference light curve,
as already noted by Schechter et al. (2003).
Figure~\ref{LC1mLC2_Res} shows the difference
between the $A$ and $B$ light curves after applying
the best-fit linear trend and time delay.
In order to subtract the two light curves we interpolated
the time-delay- and linear-trend-corrected
light curve of image $B$ to the times of image-$A$ observations.
Note that this plot mixes the $V$- and $R$-band measurements.
The error bars were calculated according to:
\begin{equation}
\Delta{m}=\sqrt{\epsilon_{A}^{2} + \epsilon_{Bn}^{2} + sf(\Delta{t}_{min})},
\label{Res_Error}
\end{equation}
where $\epsilon_{A}$ is the error for image $A$, and
$\epsilon_{Bn}$ is the error for image $B$ at the point nearest to
the interpolation time.
The structure function of the residual light curve, shown
in Figure~\ref{SF_LC1mLC2_Res}, rises rapidly from $0$ to $50$~days lag
(time in the observer system)
and then stays approximately constant
at a level of $5\times10^{-3}$~mag$^{2}$ ($\approx0.07$~mag).
The structure function was calculated using Eq.~\ref{sf},
where the DCF was calculated with $14$~days bins.

The short-timescale, uncorrelated, variability in the light curves
is a second source of systematic error in the delay measurement.
To evaluate the effect of these fluctuations on the delay accuracy,
and to test the hypothesis that the time delay
might be completely different (e.g., Gil-Merino et al. 2002), we have performed
additional Monte-Carlo simulations.
We have used the algorithm of Timmer \& Koenig (1995) to
artificially generate random red-noise light curves.
The light curves were generated with the same temporal sampling
pattern as the real observations obtained by OGLE and by us,
and with a power density spectrum (PDS) proportional to
${f}^{-\alpha}$, where $f$ is the frequency
and $\alpha$ is the power-law index.
The PDS of quasar variability in the optical range is poorly known
(see Markowitz et al. 2002, for the PDS in X-rays)
but is likely a power law with $\alpha$ between $1$ and $2$ (Giveon et al. 1999).
The standard deviation (StD) of the
variations about the mean of the
simulated light curve was
scaled to the StD of the
observed image-$B$ light curve.
To simulate the image-$A$ light curve,
we duplicated the image-$B$ light curve with a time delay $\tau_{S}$.
We then added to each point in both simulated light curves a normally
distributed noise with StD taken from
the individual errors in the observed light curves.
In order to mimic the uncorrelated noise,
we also added, in some of the simulations, the residual light curve
(shown in Fig.~\ref{LC1mLC2_Res}) to the simulated image-$A$ light curve.
We then used $\chi^{2}$ minimization, as for the real data,
to search for the best time delay.
For each set of parameters (e.g., $\alpha$, $\tau_{S}$)
we repeated this process $300$ times.
Note that we do not take into account the linear trend
in the simulations.

Our main conclusions from the simulations are:
(i) The power-law index $\alpha$ has the largest effect on the
uncertainty of the deduced time delay,
with a $\sim\pm1$~day $95\%$-confidence error for $\alpha=2$,
and a $\sim\pm7$~day $95\%$-confidence error for $\alpha=1$.
A small value of $\alpha$
introduces variability on timescales shorter than the mean
sampling interval;
(ii) When the uncorrelated variability is added to the
simulated light curve of image $A$,
the confidence interval increases by about $30\%$;
(iii) Testing for $\tau_{s}\neq-161$~day (e.g., $\tau_{s}\approx-300$~day;
as suggested by Gil-Merino et al. 2002) gives similar uncertainties
in the measured time delay to those found using $\tau_{s}=-161$~day.
Combining the uncertainty from
these simulations ($\alpha=1$), and the systematic errors described above,
we have adopted a time delay of $-161_{-7,-11}^{+7,+34}$~day ($68\%$ and $95\%$-confidence errors).
Moreover, we can reject with high confidence the possibility that
the time delay is in the vicinity of $\sim-300$~day, as suggested by earlier works on HE1104$-$1805.

To verify our result using another method, we have applied the ZDCF
cross-correlation technique (Alexander 1997) directly to the data.
We find a best fit time delay of about $-156_{-15}^{+5}$~day,
with a peak correlation of $0.82\pm0.03$,
consistent with our $\chi^{2}$ fit.
The result changes by less than $2$~days
if, before cross-correlating, we subtract the linear trend ($S$) from the image-$A$
light curve, subtract first degree polynomials from both
light curve, or use the image fluxes instead of magnitudes.


\section{Discussion and Summary}
\label{Summary}

We now discuss briefly the implications of the time delay we have found,
$\tau=-161\pm7$~days.
Leh{\' a}r et al. (2000) modeled the lensing potential of
HE1104$-$1805 using a singular isothermal
ellipsoid or a constant mass to light ratio, with and without the external shear
expected from galaxies projected within $\sim20''$, assuming they are
at the lens redshift.
Their predicted time delays for these four different models
are in the range of $-129h^{-1}$ to $-263h^{-1}$~days.
Assuming $h=0.7$ (Bennett et al. 2003),
the shortest time delay ($-184$~days) predicted by
the singular isothermal ellipse + external shear model
is larger than our result.
Note that HE1104$-$1805 is a double-image lensed quasar,
and therefore it was necessary to use 
the flux ratio between images
in order to constrain the lens model.
Interestingly, Leh{\' a}r et al. (2000) note that an external
shear twice as large as their most extreme model is needed in order for the
lens mass to be aligned with its light.
A larger shear would lower the predicted time delay,
and make it more consistent with our measurement and
with current measurement of $H_{0}$.
Alternatively, if we adopt the model range of Leh{\' a}r et al. (2000),
our measured limit of $\tau>-172$~day ($95\%$ CL) sets
a lower limit of $h>0.75$.
HE1104$-$1805 apparently is an unusual system, as
indicated by the fact that the bright image $A$ is
the one closer to the lens and that the measured
time delay is not consistent with the time delay predicted by simple models.
We defer dealing in more detail with the
implications of the time delay to a future work.

The mean flux ratio of $A/B\sim4.4$ we find
in this work is significantly different from
the emission line flux ratio of $A/B\sim2.8$
reported by Wisotzki et al. (1993).
Extrapolating the linear trend we have found,
$S\sim0.04$~mag~yr$^{-1}$, into the future,
suggests that the broad-band flux ratio between the images
will decrease to the level of the
emission-line flux ratio ($A/B\sim2.8$)
in about a decade.
At the source and lens redshifts of HE1104$-$1805,
the Einstein-radius crossing time for stellar objects
in the lens galaxy, having mass $M$
and transverse velocity $v$ is
$20.3 (M/M_{\odot})^{1/2} (v/600{\rm km~s}^{-1})^{-1} (h/0.7)^{-1/2}$~yr.
Thus, the slow trend is well explained by microlensing
in the macro images of HE1104$-$1805.

As shown in Figures~\ref{ShiftedLC} and \ref{LC1mLC2_Res},
even after removing a longterm linear trend,
there is significant uncorrelated variability.
As already noted by Schechter et al. (2003),
judging from the amplitude and timescale of the variability,
it seems that most of the uncorrelated variability
occurs in image $A$ (the one nearest to the lens).
Uncorrelated short-timescale
variability has been observed in
other lensed system (e.g., Burud et al. 2002).
However, in HE1104$-$1805 it has larger amplitudes,
with a mean of $0.07$~mag
(up to $30\%$ peak-to-peak) on timescales of less than a month.

Wambsganss et al. (1990) have
pointed out that, for large microlensing optical depth,
some events can have timescales that are considerably shorter
than the Einstein-radius crossing time.
This is caused, for example, by the passage of the source near a cusp.
Wyithe \& Loeb (2002) have suggested that the low-amplitude, fast ($\sim50$~days),
uncorrelated variability observed in RXJ0911+05 (Hjorth et al. 2002)
and in SBS1520+530 (Burud et al. 2002)
can be explained by stellar microlensing of a smooth accretion disk
that is occulted by optically thick broad-line clouds.
Although their model predicts the short timescale variability,
it cannot reproduce the large amplitude
observed in HE1104$-$1805.
Another possibility they consider is microlensing
by planetary-sized objects of $10^{-2}-10^{-4}$~M$_{\odot}$,
but they show that this cannot produce
variations with $10\%$ amplitude or greater.
Gould \& Miralda-Escude (1997) have suggested that the
microlensing of hot spots (or any structure) in the
fastly rotating quasar accretion disk
can give rise to fast variations with timescales
of $\sim1$~month.
Wyithe \& Loeb (2002) have simulated microlensing of
an accretion disk with $100$ spots.
They find a typical variability amplitude of $\sim10\%$
on one-month timescales.
Schechter et al. (2003) have argued that the variability
in HE1104$-$1805 is best explained by such a model
with $v/c\sim0.25$.
In order to test these models,
we currently continue to observe this system frequently in the $VRI$ bands.
With more data at hand,
we will attempt to re-address
the nature of the microlensing in this system in a future paper.

To summarize our main results,
we have measured the time delay of the lensed double-image
quasar HE1104$-$1805.
We have used detailed simulations at every stage of the reduction
and analysis in order to obtain
realistic error estimates.
We find that the light curves are best fit with a time delay of
$-161\pm7$~days, and with a linear trend between the images of about
$\sim0.04$~mag~yr$^{-1}$.
Our measurements resolve the previous ambiguities pertaining
to the time delay in this system.
However, the time delay is shorter than predicted
by any existing models.
The linear trend is likely due to stellar-mass microlensing
in the lens galaxy.
We confirm previous reports that
the residual light curve between the linear-trend-corrected
and time-delay-shifted light curves shows significant variability on short timescales
of about one~month.
Multi-band and spectroscopic observations
could help explain the nature of this fast variability.

\acknowledgments
We thank Paul Schechter, Lutz Wisotzki, Avishay Gal-Yam, Ohad Shemmer, Shay Zucker
and Orly Gnat for valuable discussions,
and an anonymous referee for an extremely prompt review.
This work was supported by a grant from the
German Israeli Foundation for Scientific Research and Development.

\end{document}